\documentclass[onecolumn,prd,amsmath,amssymb,floatfix]{revtex4}

\usepackage{amsmath,amssymb}
\usepackage{bm}
\usepackage{epsfig}

\newcommand{\bpsi}{\bar{\psi}}

\usepackage{slashed}

\begin{document}

\title{ 
SU(2) low energy quark effective  couplings  in weak  external magnetic field
}

\author{ F\'abio L. Braghin }
% $^{1,2}$

\affiliation{$^1$ Instituto de F\'\i sica, Fed. Univ. of  Goias, P.B.131, Campus II, 
74001-970, Goi\^ania, GO, Brazil
}

\begin{abstract}
In this work  corrections to the usual flavor SU(2) Nambu-Jona-Lasinio coupling
 due to   a weak  external magnetic field
 are calculated by considering quark polarization in a 
 (dressed)  gluon exchange mechanism for quark interactions.
The quark field is split into two components,
one that condenses and another one that  is a background field for
 interacting  quarks, being   the former  integrated out.
The resulting  determinant is expanded for 
 relatively large quark mass and small  magnetic field,
$(eB_0/{M^*}^2)<1$
by resolving 
 magnetic field dependent   low energy  quark effective interactions.
Besides the corrections for the NJL and vector NJL effective couplings,
different $B_0-$dependent 
effective couplings that break isospin and  chiral symmetry 
emerge.
\end{abstract}

\maketitle

% 12.40.Yx	Hadron mass models and calculations
% 13.40.Ks	Electromagnetic corrections to strong- and weak-interaction processes

PAC: 13.40.Ks, 
 12.39.Fe, 12.38.Lg, 12.39.Ki,  14.65.Bt

\section{
Introduction 
}

Quark interactions   
involve a large variety of different effects and mechanisms.
To identify their particular roles in observables
 and to establish a realistic  hierarchy among all of  them
for each of the processes  under their conditions within
the complexity of Quantum Chromodynamics (QCD) is a  difficult task.
 High energy density (temperature and baryon density)
 systems are known to be suitable 
to test quark
(and gluon) dynamics,  from relativistic (heavy) ion collisions (r.h.i.c.)
 to several dense stars.
Magnetic 
fields are also expected to be sizeable in such systems
\cite{reviews-B-qcd,review-B-general,tuchin}
and,  actually,  they are
expected to produce a large variety of 
  effects not only in such high energy systems 
but also in the vacuum  by means of phenomena such 
as the magnetic catalysis and the inverse effect  at finite temperatures, for example
Refs  \cite{review-B-general,magn-catalysis,Meff-B,
chiral-cond,mass-generation,magnetic-inibition,mag-inibition-2},
to produce changes  in the 
CP violation phase transition
 \cite{CP-vio-B},
the  emergence of superconducting vacuum
\cite{chernodub}
or 
 chiral asymmetry/ imballance and the chiral magnetic effect
\cite{ch-imbal-1,cme-1,ch-asym-1}
 among others.
In particular, 
it has been argued that  finite temperature inverse magnetic catalysis may be
traced back to the chirality imbalance
\cite{ch-imb-invmagcat}.
In the core of dense stars (magnetars) 
and in the early Universe
magnetic fields  are expected to be of the order of $e B_0 \sim 10^{15}$G
and 
 in non central  r.h.i.c. 
 they  may reach  
$e B_0 \sim 10^{18}$ G $\sim m_{\pi}^2$ 
or  $0.04-0.3$GeV$^2$  
from RHIC to LHC
\cite{reviews-B-qcd,tuchin,quarkmagnet},
even if within a short  time interval.
More recently it has been envisaged that one of 
the most emblematic quark-quark effective interactions,
the NJL coupling 
\cite{NJL,NJL2},  might receive a magnetic field contribution
due to   the  QCD coupling constant  dependence  on $B_0$
 \cite{G-B-2,krein-etal,G-B,ayala-g2,G-B-3,alpha-B}, 
being that for strong $B_0$ an explicit  form for the 
corrected running coupling constant
has been  derived \cite{magn-catalysis,review-B-general}.
Anisotropic contributions for the NJL coupling
 have also been found 
\cite{G-B-2,anisotropic}.
Although the usual benchmark for the investigation
of low energy effects of quark dynamics  in a magnetic field,
including   dynamical chiral  symmetry breaking, 
is the NJL model, other hadron models can also be considered 
and compared \cite{hadron-models,B-LSM}.
Besides that, 
it has been shown that vector NJL interaction
provides meaningful corrections for quark dynamics and Strong 
Interactions phase diagram  \cite{vNJL,vNJL-Bfield,vNJL2}.
 If   quark NJL and vector NJL  interactions
 receive 
corrections due to magnetic fields
 they might  produce  relevant effects in 
quark dynamics favoring or not  chiral imbalance or vector condensation.

Even before the establishment of quantum electrodynamics, 
 vacuum fluctuations for the electromagnetic field had already
been
 calculated  with the  Euler-Heisenberg action \cite{EH1936}.
With  QED 
several approaches have been employed to 
derive effective actions or Hamiltonians for higher order
contributions of 
the electromagnetic field firstly  in the absence and then in the presence
 of matter, few examples are  given  in  Refs.
\cite{schwinger51,nambu-seff,ozaki-2014}.
For several strongly interacting systems where magnetic fields
are sizeable and relevant
 it becomes interesting to investigate the vacuum
 polarization effects
in the presence of magnetic fields.
In this work,  effective quark-quark interactions are calculated
 in the presence of constant weak magnetic field from vacuum polarization
effects.
The one loop
background field method for quarks, 
as employed in Refs. \cite{PRD2014,piquark,higher-orders}, will be considered
in the presence of a constant 
weak magnetic field.

The departure point of the present work 
is the global color model (GCM) 
 obtained 
by considering  gluon exchange  corrected by 
 gluon interactions and its non-Abelian character, i.e. it can be a realistic gluon propagator.
It is given by \cite{ERV,PRC,bossche}:
\begin{eqnarray} \label{Seff}  
S_{eff}  [\bar{\psi}, \psi] = \int_x \left[
\bar{\psi} \left( i \slashed{\partial} 
- M \right) \psi 
- 
 \frac{g^2}{2}\int_y j_{\mu}^b (x) 
(R^{\mu \nu}_{bc} )^{-1} (x-y) j_{\nu}^{c} (y) 
 \right]
,
\end{eqnarray}
Where 
the color  quark current is $j^{\mu}_a = \bar{\psi} \lambda_a \gamma^{\mu} \psi$,
 the sums in color, flavor and Dirac indices are implicit,
the kernel ${(R^{\mu \nu}_{bc})}^{-1}$ is the gluon propagator.
  Non abelian 
gluon interactions  
 can be considered 
 to dress the gluon exchange
by considering a non perturbative (realistic)  gluon propagator
that, together with the quark-gluon  vertex,
will be assumed to provide the  strength 
for Dynamical Chiral Symmetry Breaking  (DChSB). 
Several different effects are known to contribute
to the strength of the quark-quark  interaction  above
 \cite{SD-rainbow,kondo,cornwall}.
Therefore this calculation 
presents, in this  sense, a similar level of approximation  to 
 the rainbow ladder approximation 
for the Schwinger Dyson equations 
that  yield
DChSB \cite{PRC,ERV,bossche,SD-rainbow,maris-craig,higa,aoki}.
% \cite{SD-rainbow}.
The quark gluon vertex  was shown to depend on $B_0$ \cite{ayala-g2}
and this will not be considered in the present work.
This model will be coupled to the electromagnetic 
field via the quark minimal coupling.
To investigate the flavor structure of the model, 
one  performs a Fierz transformation 
 from which
a NJL coupling emerges in the local limit,
  besides other structures.
This work is organized as follows.
In the next Section
the Fierz transformation of this GCM interaction coupled (minimally) 
to  a constant magnetic  field  is presented and the 
quark field is integrated out  in the presence of  background 
quark.
In the following Section the determinant is expanded for 
small magnetic field with respect to the  quark effective mass,
i.e. $(e B_0) < < {M^*}^2$.
Several simple ratios between the effective couplings in the limit 
of large  quark effective mass are obtained.
 In the final Section a summary and discussion  are presented.

\section{
Quark components and light  meson fields
}

The generating functional to be considered is the following:
$$
Z [ \xi, \bar{\xi}] = N  \int {\cal D}[\bpsi,\psi]
e^{ i \int ( {\cal L} +\bpsi J + \bar{J} \psi) },
$$
where the Lagrangian density for the 
 minimal electromagnetic coupling (for a background electromagnetic field)
to the global color model
 can be  written as:
 \begin{eqnarray} \label{Seff}  
{\cal L}  = 
\bar{\psi} \left( i \gamma \cdot { D}
- m \right) \psi 
-
\frac{g^2}{2}
 \int_y \bpsi (x)\gamma_{\mu} \lambda^b \psi (x)
 R^{\mu \nu}_{bc}  (x-y)\bpsi(y) \gamma_{\nu} \lambda^c \psi (y)
,
\end{eqnarray}
Where 
$a,b...=1,...(N_c^2-1)$ stand 
for color in the adjoint representation, 
and $i,j,k=0,...(N_f^2-1)$ will be used  for SU(2) flavor indices,
 the sums in color, flavor and Dirac indices are implicit,
and the covariant quark derivative  is:
$D = D_{\mu} = \partial_\mu \delta_{ij} - i e Q_{ij} A_{\mu}$
for the diagonal matrix $\hat{Q}
=  diag(2/3, -1/3)$.
In several  gauges, the gluon kernel
can  be  written in terms of the transversal and longitudinal components
as:
 $R^{\mu\nu}_{ab} (x-y) = \delta_{ab} \left[
 (g^{\mu\nu} - \frac{\partial^\mu \partial^\nu}{\partial^2})  R_T (x-y)
+ \frac{\partial^\mu \partial^\nu}{\partial^2}
R_L (x-y) \right]$.
 The infrared regime of the gluon propagator exhibits
a non trivial behavior that is often parameterized in terms of 
an effective gluon mass \cite{gluonmass}.
This will be discussed further in Section IV.

To make possible a more detailed  investigation of the different 
flavor channels of quark 
interactions a 
Fierz transformation  \cite{NJL2,IZ}
 can be done next.
Then, for  the quark interaction above: 
\begin{eqnarray}
\Omega \equiv
\frac{g^2}{2} 
\bpsi (x)\gamma_{\mu} \lambda^b \psi (x)
 R^{\mu \nu}_{bc}  (x-y)
\bpsi(y) \gamma_{\nu} \lambda^c \psi (y),
\nonumber
\end{eqnarray}
 the Fierz transformed 
${\cal F} (\Omega)$ color singlet expression is given by:
\begin{eqnarray} \label{fierz4}
{\cal F} (\Omega)&=&
 \alpha g^2
 \left\{ 
\left[ 
j_S (x,y)  j_S(y,x) +
 j_P^i(x,y)  j_P^i(y,x)  
\right] 
R (x-y) \right.
\nonumber
\\
&-&  \left. \frac{1}{2} \left[ j_{V,\mu}^i (x,y) j_{V,\nu}^i (y,x) 
-   {j_{\mu}^i}_A (x,y)  {j_{\nu}^i}_A (y,x)
\right] 
 \bar{R}^{\mu\nu}  (x-y)
\right\} ,
\nonumber
\end{eqnarray}
where 
$\alpha = 
 8/9$ for  SU(2)  flavor,
and the 
following  bi-local quark bilinears 
($j_i^q (x,y) =  \bpsi (x) \Gamma^q \psi (y)$ 
for operators $\Gamma_q$  where $q=s,p,v,a$)
 were defined:
\begin{eqnarray}
j^S(x,y) &=&  \bpsi (x) \psi (y), \;\;\;\;\;\;\;\;\;\;\;\;
 j_i^P(x,y) =  \bpsi (x)  \sigma_i i \gamma_5 \psi (y),
\nonumber
\\
j_i^{V,\mu}(x,y) &=&  \bpsi (x) \gamma^\mu \sigma_i \psi (y),
 \;\;\;\;\; j_i^{\mu,A}(x,y) = 
\bpsi (x)   i \gamma_5 \gamma^\mu \sigma_i \psi (y),
\end{eqnarray}
In these expressions
 following kernels were used:
\begin{eqnarray}
R (x-y) &\equiv&  R = 
3 R_T (x-y) + R_L (x-y),
\nonumber
\\
\bar{R}^{\mu\nu} (x-y) &\equiv& \bar{R}^{\mu\nu} 
= g^{\mu\nu} (R_T (x-y) +R_L (x-y) ) + 
2 \frac{\partial^{\mu} \partial^{\nu}}{\partial^2} 
(R_T (x-y)  - R_L (x-y) ).
\end{eqnarray}
The long-wavelength or  local limit of  the scalar and pseudo-scalar   interactions 
 yields the
% Nambu Jona Lasinio (NJL)
NJL coupling
 with 
 $G \sim \frac{g^2}{\Lambda_{qcd}^2}$ 
or $G \sim \frac{g^2}{M_G^2}$ for massless and massive  gluons
\cite{A-E,PRD2014,1002.4151}.

The quark field will be split according to the Background Field Method (BFM)
\cite{background,SWbook}.
One component is considered to be a
(constituent quark)  background field ($\psi_1$), 
 and the sea quark field ($\psi_2$) will be integrated out.
This splitting of the field can be made by means of the bilinears 
 $\bpsi \Gamma \psi$
where $\Gamma$ stands for Dirac, color  or flavor operators,
such that the resulting determinant corresponds basically to the 
one loop  BFM results.
The splitting can be written as
\cite{PRD2014,piquark}:
\begin{eqnarray} \label{split-Q} 
\bpsi \Gamma^q \psi &\to& (\bpsi \Gamma^q \psi)_2 + (\bpsi \Gamma^q \psi)_1,
\end{eqnarray}
where  $(\bpsi \psi)_2$ will be integrated out,
being possible that it 
 composes  light mesons and the scalar condensate 
and  the  component $(\bpsi \psi)_1$ stands for the
 background field that yields   baryon constituent 
quarks.
This separation 
   preserves chiral symmetry,
and it may  not correspond to  a simply mode separation  of low and high energies 
which might be a very restrictive assumption since pions and constituent quarks might 
be composed by   quarks with similar energy  modes (fully or in part). 
Therefore it seems the criterium might not
involve a  separation  scale  
and, at the end,  the shape of the results should be  basically  the same.
The shift of bilinear may also be suitable for envisaging quark-anti-quark
states which are the most important states for the very low energy 
QCD, i.e. below the nucleon mass scale.
The effective interaction $\Omega$ is split accordingly
and terms with mixed bilinear  of $\psi_1$ and $\psi_2$ can be written such that
the quadratic part of bilinear $\bar\psi_2 \psi_2$ will be suitable to be integrated out.
The interaction $\Omega_2$
deserves some more attention and it can be handled in  two ways:
(1)  By resorting to  a weak field approximation $\Omega_2 < < \Omega_1$
which yields directly the one loop BFM that might receive corrections by 
a perturbative expansion which incorporates $\Omega_2$ \cite{background};
(2)  By making use of the 
auxiliary field method according to which
a set of auxiliary fields  is introduced by means of unitary functional integrals
multiplying the generating functional
 \cite{ERV,PRC,piquark,kleinert,EPJP}.
Auxiliary fields (a.f.) 
 allow to incorporate properly  DChSB with the formation of the 
scalar quark condensate which endows quarks with a large effective mass.
Therefore the use of the a.f.  improves  the one loop
background field method as usually implemented.
Auxiliary fields are introduced by multiplying the generating functional 
by the following normalized Gaussian integrals:
\begin{eqnarray}
 1 &=& N \int D[S] D[P_i]
 e^{- \frac{i}{2 } t_2^2
\int_{x,y}   R  \alpha  \left[ (S - g   j^S_{(2)})^2 +
(P_i -  g    j^{P}_{i,(2)} )^2 \right]}
\nonumber
\\
&&
 \int 
D[V_\mu^i]
\int  D[\bar{A}_\mu^i] 
 e^{- \frac{i}{4 } t_2^2
\int_{x,y} {\bar{R}^{\mu\nu}} \alpha
\left[ (V^i_{\mu} -  g    j_{V,\mu}^{i,(2)}) (V^i_{\nu} -
 g   j_{V,\nu} ^{i,(2)} )
\right] }
 e^{- \frac{i}{4 } t_2^2
\int_{x,y} {\bar{R}^{\mu\nu}} \alpha
\left[ 
  ( \bar{A}^i_{\mu} -  g    {j_\mu^{i,(2)}}^A ) ( \bar{A}^i_{\nu} -
 g   {j_\nu^{i,(2)}}^A )
\right]
}.
\end{eqnarray}
In these  expressions
the bilocal a.f. are $
S(x,y), P_i(x,y), V_\mu^i(x,y)$ and $\bar{A}_\mu^i(x,y)$ 
and they have been shifted by quark currents such as to cancel out the fourth 
order interactions $\Omega_2$.
These shifts have unit Jacobian and they generate a coupling to quarks.
The non locality of these auxiliary fields give rise to form factors 
 which nevertheless  can produce punctual  meson fields
by expanding in an infinite basis of local fields.
Finally it is also possible to consider the long-wavelength limit
by  keeping 
only the lowest energy states and by simply considering
 the local limit for structureless light mesons  \cite{piquark}.
The resulting 
effective action  for quarks  ($\psi_1$ and $\psi_2$) interacting with 
auxiliary fields (quark-anti-quark mesons)  is quadratic in  $\psi_2$
requiring a typical Gaussian integration.
The resulting determinant can be written, 
by considering the identity
$\det A = \exp \; Tr \; \ln (A)$, 
as:
\begin{eqnarray} \label{Seff-det}  
S_{eff}   &=&    i  \; Tr  \; \ln \; \left\{
{S_0}^{-1} (x-y) 
+ \Xi (x-y)
\right.
\nonumber
\\
&-& \left.
\alpha g^2 \bar{R}^{\mu\nu} (x-y) \gamma_\mu  \sigma_i \left[
 (\bpsi (y) \gamma_\nu  \sigma_i \psi(x))_1
+  i \gamma_5   (\bpsi  (y)
i \gamma_5 \gamma_\nu  \sigma_i \psi (x))_1 \right]
\right.
\nonumber
\\
&+& 
\left.
 2   R (x-y) \alpha g^2
 \left[  (\bpsi (y) \psi(x))_1 
+ i  \gamma_5 \sigma_i  (\bpsi (y) i \gamma_5  \sigma_i \psi (x))_1 \right]
 \right\} 
\nonumber
\\
&-& \frac{1}{2 }
\int_{x,y}  \left\{ R    \left[ S^2 +
P_i^2 \right]
+ \frac{1}{2 }
 {\bar{R}^{\mu\nu}} 
\left[ V^i_{\mu} V^i_{\nu}  +
\bar{A}^i_{\mu}  \bar{A}^i_{\nu} 
\right] \right\},
\nonumber
\\
&-&  
\int_{x}
 \bar{\psi}_1 (x) \left( i \gamma_{\mu} { D}^{\mu} 
- m \right)\psi_1 (x)
- \frac{g^2}{2} \int_{x,y}
 j_{\mu}^{a,(1)} (x) R^{\mu\nu}_{ab} (x-y) j_{\nu}^{b,(1)}(y)
\end{eqnarray}
where 
$Tr$ stands for traces of discrete internal indices 
and integration of  spacetime coordinates,
 the inverse Fierz transformation was done
for the  $\psi_1$   interaction that is written  in the last line,
and where 
$S_0^{-1} = \left(  i \slashed{D} -  m
\right)$,
being 
 $\slashed{D} =  \gamma^\mu( \partial_\mu \delta_{ij} - i e Q_{ij} A_{\mu} )$.
$\Xi$ 
stands for
 the auxiliary fields coupling to sea quarks.
 Vector and axial  auxiliary fields yield  heavier
excitations and may be neglected for  the low energy regime. 
The bilocal 
a.f.  can be expanded in a basis of local meson excitations.
However,  this work is concerned with the effects of weak magnetic field in the low energy regime
of quark effective interactions and
the local limit of these composite fields might  be adopted  
because  the only  leading effect
of the a.f. is to produce the large quark effective mass due to DChSB.
The quark coupling to the local scalar and pseudo-scalar fields,
 in the absence of the heavier vector 
states,  
can be written as:
\begin{eqnarray} \label{q-meson-term}
\Xi (x,y) =
g \alpha 
 F_{0,0} (x-y)  R  \left[ S \left(\frac{x+y}{2} \right)  + 
P_i  \left( \frac{x+y}{2}\right) i \gamma_5 \sigma_i   \right] 
\end{eqnarray}
where, 
 due to the structureless mesons approximation,  it will be considered  $z=(x+y)/2=x$.
Then it
reduces to:
\begin{eqnarray} \label{scalars-q}
\Xi (x,y)   \simeq  F 
  \left[s (x) + p_i (x) \gamma_5 \sigma_i  
\right] \delta (x-y) ,
\end{eqnarray}
where $F$ is the pion decay constant that allows for the canonical definition of the pion field
as $\pi_i = F p_i$.
The saddle point equations for expression (\ref{Seff-det}) yield the usual gap equations,
 by denoting the auxiliary fields $\phi_q=S(x,y),P_i(x,y),
V^\mu_i (x,y)$ and the axial field $\bar{A}^\mu_i (x,y)$ these equations are:
\begin{eqnarray} \label{gap-eq}
\frac{\partial S_{eff}}{\partial \phi_q} = 0.
\end{eqnarray}
These equations for the NJL model and GCM  have been analyzed in many works,
under external B or not, for the vacuum or at finite temperatures or quark densities,
including in the complete form which corresponds to Dyson Schwinger equations 
in the rainbow ladder approximation.
 The only possible non trivial solution might exist for the scalar field since the
ground state is scalar. 
It yields a correction to the 
quark mass, as the leading effect, and therefore 
only the limit of local auxiliary field $S$ is needed from here on.
The magnetic field is known to increase the effective mass in the 
magnetic catalysis effect, for example in Refs. \cite{Meff-B,chiral-cond,cond-B,review-B-general}.
By considering solutions for which 
 the quark-gluon coupling of the model is sufficiently strong to generate
DChSB, as shown in Section IV,  it yields  a correction to the 
 quark effective mass  ($M^*$) such that the quark kernel in expression (\ref{Seff-det})
 receives 
a correction, being written as:
\begin{eqnarray}
% S_0^{-1} \to 
S_0^{-1}  = \left(  i \slashed{D} -  M^*
\right),
\end{eqnarray}
 At this point  it is worth noticing that an estimate of the effect of the a.f. on the 
eventual quark-quark effective interactions can be obtained by expanding the quark determinant
above in powers of the a.f. 
However it is seen that the effects of a.f. on the quark-quark  effective interactions
only will appear at least in the third order of the expansion  to produce, for example,
terms of the following form
$\phi_q (\bpsi \Gamma_q \psi)^2$.
These terms   are of higher order in $S_0^n$ and consequently numerically 
smaller in the large quark mass limit. 
 Alternatively,
if these auxiliary fields are kept as a whole and afterwards 
eliminated  for example 
  being integrated out approximately when  expanding the determinant
in a steepest descent  approximation, their 
contribution to the photon-quark interaction would  be again 
of higher order.
Therefore the a.f. can be neglected by keeping the non trivial value of the scalar field
that endows quarks with an effective mass.
Results will be precisely those  from  one loop BFM with 
the corrected quark effective mass.

The determinant can be rewritten as:
\begin{eqnarray}
I_{det} &=&  Tr \ln \left( S_0^{-1} + \sum_q a_q \Gamma_q j_q \right)
\nonumber
\\
\label{Idet}
&=&
\frac{1}{2} Tr \ln 
\left[  \left( S_0^{-1} + \sum_q a_q \Gamma_q j_q \right)
 \left( \bar{S}_0^{-1} + \left(\sum_q \bar{a}_q \Gamma_q j_q \right)^* 
 \right)
\right]
\end{eqnarray}
where 
$\bar{S}_0^{-1} = 
\left( 
 i \slashed{D} +  M^*
\right)$; 
for $q=s,p,v,a$ and   $\bar{a}_s=-a_s$, $\bar{a}_v=-a_v$ and 
 $\bar{a}_p= a_p$, $\bar{a}_a=a_a$,
and also 
 it has been defined the following shorthand notation for the four channels $q$:
\begin{eqnarray}
 \sum_q  a_q \Gamma_q j_q &=&
-
\alpha g^2 \bar{R}^{\mu\nu} (x-y) \gamma_\mu  \sigma_i \left[
 (\bpsi (y) \gamma_\nu  \sigma_i \psi(x))
+  i \gamma_5   (\bpsi  (y)
i \gamma_5 \gamma_\nu  \sigma_i \psi (x)) \right]
\\
&+& 
 2  \alpha g^2   R (x-y) 
 \left[  (\bpsi (y) \psi(x))
+ i  \gamma_5 \sigma_i  (\bpsi (y) i \gamma_5  \sigma_i \psi (x)) \right].
\nonumber
\end{eqnarray}
 By turning the (background) quark currents
to zero this determinant   yields the celebrated Euler Heisenberg effective action 
for the electromagnetic field 
\cite{EH1936,schwinger51,IZ,cond-B}.
Below,  a large quark mass expansion will be performed
and  the leading
quark-quark effective  
 couplings and their dependence on 
a constant magnetic field will be shown.

\section{ Expansion of the determinant and effective couplings}

The large quark mass expansion of the determinant will be performed next 
by neglecting all the quark derivative couplings \cite{mosel}.
A shorthand notation will be used below 
to improve the reading of the expressions, the gluon kernels
will be written shortly:
$R\equiv R(x-y)$, $\bar{R}^{\mu\nu}\equiv  \bar{R}^{\mu\nu}(x-y)$ and so on.
By neglecting terms such as $Tr \ln (i S^{-1}_0)$ that 
becomes an irrelevant constant in the generating functional,
the  dynamical part of expression
(\ref{Idet}), 
 by considering the anti-commutation relations
of the Dirac gamma matrices,  
can then  be written as:
\begin{eqnarray}   \label{det-exp}
S_{d} &\simeq& 
Tr \sum_{n=1}^\infty
d_n 
%&& 
\left\{   
\tilde{S}_2
\left[ \Delta_A +  \xi
+ \xi_{sb}
+ \xi_{der}
+ I_{crossed}
+
4 (\alpha g^2)^2  R^2 \left[
j_S(x,y) j_S(y,x)
+
  \gamma_5^2     \sigma_i   \sigma_j
j_P^i(x,y) j_P^j(y,x) \right] \right.
\right.
\nonumber
\\
&-& \left.  \left.
 (\alpha g^2)^2 \bar{R}^{\mu\nu} \bar{R}^{\rho\sigma} \gamma_\mu
 \gamma_\rho \sigma_i  \sigma_j  \left[
{j_V}_\nu^i (x,y) {j_V}_\sigma^j (y,x)
- \gamma_5^2    
{j_A}_\nu^i (x,y) {j_A}_\sigma^j (y,x)
\right] 
\right]
\right\}^n,
\end{eqnarray}
where the following terms have been defined:
\begin{eqnarray} \label{D-A-xi}
\Delta_A &=&
-  e^2  \hat{Q}^2 (A_\mu A^\mu + A^\mu A^\nu \sigma_{\mu\nu})
+ \frac{i e}{2} \hat{Q}
 \sigma_{\mu\nu} F^{\mu\nu}
,
\\
\xi &=&  
- e 2  (\alpha g^2) R 
\hat{Q}
 \sigma_i  i \left[ \slashed{A} \; , \; 
\gamma_5      \right]
j_P^i (x,y)
\nonumber
\\
&+& 
 e  ( \alpha g^2)  \bar{R}^{\mu\nu} 
\hat{Q}
\sigma_i 
 \left[  \left[ \gamma_\mu \; , \; \slashed{A} \right]
{j_V}_\nu^i (x,y)
+   
 \left\{ \gamma_\mu  i \gamma_5
 \; , \; \slashed{A} \right\}   {j_A}_\nu^i (x,y)
 \right] ,
\\
 \xi_{sb} &=&
4 (\alpha g^2) R M^* j_S (x,y)
- 2 M^*   (\alpha g^2) \bar{R}^{\mu\nu} \sigma_i  \gamma_\mu 
 {j_V}_\nu^i (x,y)
,
\\
\xi_{der} &=&
-
i \alpha g^2 \bar{R}^{\mu\nu} (x-y) 
\sigma_i \left[
 \left[ \gamma_\rho \; , \; \gamma_\mu \right]
\partial^\rho ( j_{V,\nu}^i (y,x))
+  i \gamma_5 
 \left\{ \gamma_\rho \; , \; \gamma_\mu \right\}
\partial^\rho   ( j_{A,\nu}^i  (y,x) ) \right] 
\nonumber
\\
&+& 
 2  \alpha g^2   R (x-y) 
 i  \sigma_i
 \left[\gamma_\rho \; , \; \gamma_5 \right]
\partial^\rho  
  ( j_P^i (y,x) ) 
,
\\
I_{crossed} &=&
 i
(\alpha g^2)^2 \bar{R}^{\mu\nu} \bar{R}^{\rho\sigma} 
\left\{ \gamma_\mu   
\gamma_\rho \gamma_5  \sigma_i  \sigma_j 
+
 \gamma_\rho \gamma_5   
 \gamma_\mu \sigma_j  \sigma_i 
\right\}
 {j_V}_\nu^i (x,y) {j_A}_\sigma^j (y,x)
\nonumber
\\
&+&
  2   (\alpha g^2)
 R 
\bar{R}^{\mu\nu}
 \left\{  
\gamma_5  \gamma_\mu \gamma_5 \sigma_i   \sigma_j 
+
  \gamma_\mu \gamma_5^2 \sigma_j  \sigma_i
 \right\}
j_P^j (x,y) {j_A}_\nu^i (y,x)
\nonumber
\\
&+&  2 i
 (\alpha g^2)
 R
\bar{R}^{\mu\nu}  \left\{ \gamma_5  \gamma_\mu  \sigma_j \sigma_i  
- 
 \gamma_\mu  \gamma_5
 \sigma_i \sigma_j 
\right\}
j_P^j (x,y)  {j_V}_\nu^i (y,x)
\nonumber
\\
&+&  2 
 (\alpha g^2)
 R
\bar{R}^{\mu\nu}   \sigma_i
 \gamma_\mu 
j_S (x,y)  {j_V}_\nu^i (y,x).
\end{eqnarray}
The terms $\Delta_A$ and $\xi$ contain  magnetic field dependent terms,
$\xi_{sb}$ presents the symmetry breaking terms since they appear to be
 proportional to the Lagrangian
quark mass $m$. 
However if  DChSB is considered for the auxiliar scalar field and 
the corresponding gap equation, this Lagrangian mass is corrected to an effective mass
$M^*$.
The other terms above are:
 $\xi_{der}$ with the  derivative terms that,
with an integration by parts,
 may produce constant magnetic field contribution 
when multiplied by 
$\xi$,
whereas  $I_{crossed}$ contains mixing  interactions 
with different quark currents and they produce
non zero terms in the expansion only in higher orders.
In this work only the lowest order terms
will be investigated, up to the second order in
the expansion.
Third order of the expansion will have additional factors 
$\tilde{S}_2$, being each of them ${\cal O}(1/{M^*}^{2})$ smaller
than the second order ones.
In  expression (\ref{det-exp})
the following parameters were defined:
\begin{eqnarray} 
d_n &=& -  i \frac{(-1)^{n+1} }{2 n},
\\
\tilde{S}_2 &=& 1/(- \partial^2 - {M^*}^2)
.
\end{eqnarray}

\subsection{First order terms}

In the long-wavelength or  local limit of the expressions bellow
the  effective couplings can be resolved to yield  
effective coupling constants.
In the zero order derivative expansion for the first order expansion
the following effective couplings appear:
\begin{eqnarray} \label{4quark-B}
{\cal I}_{eff,1} &=&  
\;  
\Delta M^* \bpsi \psi
+   g_{4}   \left[ ( \bpsi  \psi )^2
 +   (  \bpsi \sigma_i  i \gamma_5 \psi )^2  \right]
-  
\;
 g_{4v}  
\left[ ( \bpsi  \sigma_i \gamma_\mu \psi )^2
 + ( \bpsi \sigma_i \gamma_\mu \gamma_5 \psi )^2
\right] .
\end{eqnarray}
These couplings have already been found in Refs.  \cite{piquark,higher-orders} and,
for these expressions,   traces of Dirac and Pauli matrices are taken.
The effective coupling constants were defined    in the following way:
\begin{eqnarray}  \label{NJL-couplings}
\Delta M^* 
&=& 
-  i 2 (\alpha g^2)  
\; Tr \; \left( \tilde{S}_2  R
\; 
 M^* \right)
,
\nonumber
\\
g_{4}  ( 1 \; , \; \delta_{ij} )
&=&
 -  i  2 (\alpha g^2)^2    
\;   Tr \; 
\left(  \tilde{S}_2  R^2
\;
 ( 1 \; , \; \sigma_i \sigma_j )
\right),
\nonumber
\\
 g_{4v} \;  g^{\nu\sigma} \delta_{ij} 
 &=&
- \frac{ i}{2} (\alpha g^2)^2   
\;   Tr
\;  
\left( \tilde{S}_2
\bar{R}^{\mu\nu}
 \bar{R}^{\rho\sigma} \gamma_\mu
\;
 \gamma_\rho \sigma_i  \sigma_j 
\right)
 ,
\end{eqnarray}
where,  by performing the trace in Dirac indices, 
 the following kernel can be  defined:
\begin{eqnarray} \label{R2v}
\ \bar{R}_{\mu}^{\rho} \bar{R}_{\rho \nu} 
= \bar{R_{2}}_{\mu\nu}= 
g_{\mu\nu}(R_T+R_L)^2 
+ 8 \frac{k_\mu k_\nu}{k^2} R_T (R_T-R_L).
\end{eqnarray}
The expression for the effective mass (\ref{NJL-couplings}) might be 
ultraviolet divergent or finite 
depending on the gluon propagator  behavior.
However the effective couplings constants $g_4$ and $g_{4,v}$ are finite
unless the quark and gluon kernels present an unusual momentum dependence.
For gluon propagators written in terms of an effective gluon mass these expressions
should also be infrared finite.

\subsection{Second order quark terms up to ${\cal O} (\bpsi \Gamma_q \psi)^2$}

The second order
  non derivative couplings that depend  on 
 the magnetic field 
will be exhibited below.
Those terms containing  one derivative 
 of quark currents ($\xi_{der}$) 
that multiply the vector $A^\mu$
either  can 
yield non trivial contributions to (non derivative) effective quark couplings
if an integration by parts is performed, 
 producing  quark couplings to the strength tensor $F^{\mu\nu}$, 
or it may disappear.
The terms that produce non zero contributions 
are shown below  (${\cal I}_{4q} = I_2  +  I_4 + I_{4,\xi} + I_{4,der} + 
I_{cross,B}$).
The two possible orders of combining  structures  for each of the term
in the expansion   will be written
as  a big anti-commutator in most of the terms.
 Although all the calculations will be performed  for  the Landau gauge
for a constant magnetic field, $A_\mu = - B_0 (0,0,x,0)$,
the electromagnetic field will be  carried  almost until the last expressions.
These terms are  the following:
\begin{eqnarray}
I_2 &=&
i  \; e^2   (\alpha g^2) M^*
\; Tr \; \hat{Q}^2
\left\{   \tilde{S}_2
(A_\mu A^\mu ) 
\; , \;
\tilde{S}_2 R \right\}
j_S (x,y) 
, 
\nonumber
\\
I_4 &=&
- i   e^2   
 (\alpha g^2)^2
\; Tr \; \left\{   \tilde{S}_2
(A_\mu A^\mu + A^\mu A^\nu \sigma_{\mu\nu})
\; , \; 
 \tilde{S}_2 R^2
\right\}
 \left(
\hat{Q}^2
j_S(x,y) j_S(y,x)
+
  \gamma_5^2     \hat{Q}^2 \sigma_i     \sigma_j
j_P^i(x,y) j_P^j(y,x) \right)
\nonumber
\\
&-&   
\frac{i}{4}    e^2   
(\alpha g^2)^2
\; Tr \; \left\{   \tilde{S}_2
(A_\mu A^\mu + A^\mu A^\nu \sigma_{\mu\nu})
\; , \; 
\tilde{S}_2
 \bar{R}^{\rho_1\sigma_1} \bar{R}^{\rho\sigma} 
 \gamma_{\rho_1} 
\gamma_\rho 
\hat{Q}^2 \sigma_i  \sigma_j 
\right\}
\nonumber
\\
&& \times
 \left(
{j_V}_{\sigma_1}^i (x,y) {j_V}_\sigma^j (y,x)
- \gamma_5^2    
{j_A}_{\sigma_1}^i (x,y) {j_A}_\sigma^j (y,x)
\right)  %=
\nonumber
\\
&+& 
\frac{i}{8} 
i e 
(\alpha g^2)^2
\; Tr \;  \left\{
\tilde{S}_2
F^{\mu_1\nu_1}  \; , \;
\tilde{S}_2
\bar{R}^{\mu\nu} \bar{R}^{\rho\sigma} 
\sigma_{\mu_1\nu_1}
\gamma_\mu
 \gamma_\rho
\hat{Q}^2
 \sigma_i  \sigma_j 
\right\}
 \left(
{j_V}_\nu^i (x,y) {j_V}_\sigma^j (y,x)
- \gamma_5^2    
{j_A}_\nu^i (x,y) {j_A}_\sigma^j (y,x)
\right)
,
\nonumber
\\
I_{4,\xi}
&=&  
  i  e^2  (\alpha g^2)^2 
\; Tr \; \hat{Q} \sigma_i 
\hat{Q} \sigma_j
\left( 
 \tilde{S}_2 R  
 \left[ \slashed{A} 
\; , \; 
\gamma_5      \right]
\tilde{S}_2 R  
  \left[ \slashed{A} \; , \; 
\gamma_5      \right] 
\right)
j_P^i (x,y) j_P^j (y,x),
\nonumber
\\
&+&
\frac{i}{4} 
 e^2  ( \alpha g^2)^2  
\; Tr \;    \hat{Q} \sigma_i  \hat{Q}  \sigma_j  
\left( 
\tilde{S}_2 
 \left[ \gamma_\mu \; , \; \slashed{A} \right] 
\bar{R}^{\mu\nu} 
\tilde{S}_2 
 \left[ \gamma_\rho \; , \; \slashed{A} \right] 
\bar{R}^{\rho\sigma}
\right)
{j_V}_\nu^i (x,y) {j_V}_\sigma^j (y,x)  
\nonumber
\\
&+&   
\frac{i}{4} 
 e^2  ( \alpha g^2)^2  
\; Tr \;
\hat{Q} \sigma_i \hat{Q}   \sigma_j 
\left(
\tilde{S}_2
 \left\{ \gamma_\mu  i \gamma_5
% \; , \; 
\slashed{A} \right\}  
 \bar{R}^{\mu\nu} 
\tilde{S}_2
\left\{ \gamma_\rho  i \gamma_5
 \; , \; \slashed{A} \right\}  
 \bar{R}^{\rho\sigma}
\right)
  {j_A}_\nu^i (x,y) {j_A}_\sigma^j (y,x)   
\nonumber
\\
&+&
\frac{i}{4} 
 e^2  ( \alpha g^2)^2  
\; Tr \;    \hat{Q} \sigma_i  \hat{Q}  \sigma_j  
\left(
\tilde{S}_2 
 \left[ \gamma_\mu \; , \; \slashed{A} \right] 
\bar{R}^{\mu\nu} 
\tilde{S}_2 
 \left[ \gamma_\rho \; , \; \slashed{A} \right] 
(i \gamma_5)
\bar{R}^{\rho\sigma}
\right)
{j_V}_\nu^i (x,y) {j_A}_\sigma^j (y,x) 
\nonumber
\\
&+&
 i 2 e  ( \alpha g^2)^2
\; Tr \; 
M^*
\left(
\tilde{S}_2  
\hat{Q}
\sigma_i 
 \left[ \gamma_\mu \; , \; \slashed{A} \right]
 \bar{R}^{\mu\nu} 
\tilde{S}_2 R
\right) 
{j_V}_\nu^i (x,y)
 j_S (x,y)
\nonumber
\\
 &+&
i
4 (\alpha g^2)^2 
\; Tr \; 
({M^*}\tilde{S}_2 R)^2  j_S (x,y)  j_S (y,x)
\nonumber
\\
&+&  i     (\alpha g^2)^2
\; Tr \; 
\left(
 \sigma_i  \sigma_j  \gamma_\mu   \gamma_\rho 
 {M^*} \tilde{S}_2 \bar{R}^{\mu\nu}
{M^*}
 \tilde{S}_2
 \bar{R}^{\rho\sigma} 
\right) 
 {j_V}_\nu^i (x,y)  {j_V}_\sigma^j (y,x)
,
\nonumber
\\
I_{4,der} &=& 
- i^2 e   (\alpha g^2)^2  
\; Tr \; 
\left\{
\tilde{S}_2 R 
\hat{Q}
 \sigma_i   \left[ \slashed{A} \; , \; 
\gamma_5      \right]
\; , \; 
\tilde{S}_2  R
 \sigma_j
 \left[\gamma_5 \; , \; \gamma_\rho \right]
\right\}
j_P^i (x,y)
\partial^\rho  
 ( j_P^j   (y,x)  )
\nonumber
\\
&+&
\frac{i^2}{2} e  ( \alpha g^2)^2  \;
Tr \;
\left\{
 \tilde{S}_2
 \bar{R}^{\mu\nu} 
\hat{Q}
\sigma_i 
 \left(  \left[ \gamma_\mu \; , \; \slashed{A} \right]
\right)
\; , \; 
\tilde{S}_2
 \bar{R}^{\mu_2\nu_2} (x-y) 
\sigma_j
 \left[ \gamma_\rho \; , \; \gamma_{\mu_2} \right]
\right\}
{j_V}_\nu^i (x,y)
\partial^\rho (j_{V,\nu_2}^j (y,x) )
\nonumber
\\
&+&
\frac{i^2}{2}  e  ( \alpha g^2)^2 
\; Tr \; 
\left\{
\tilde{S}_2
 \bar{R}^{\mu\nu} 
\hat{Q}
\sigma_i 
 \left(  \left[ \gamma_\mu \; , \; \slashed{A} \right]
\right)
\; , \; 
\tilde{S}_2
 \bar{R}^{\mu_2\nu_2} (x-y) 
\sigma_j   \gamma_5 
 \left\{ \gamma_\rho \; , \; \gamma_{\mu_2} \right\}
\right\}
{j_V}_\nu^i (x,y)
\partial^\rho    (j_{A,\nu_2}^j (y,x) ) 
\nonumber
\\
&+&
\frac{i^2}{2}  e  ( \alpha g^2)^2  
\; 
Tr \; 
\left\{
\tilde{S}_2
\bar{R}^{\mu\nu} 
\hat{Q}
\sigma_i 
 \left\{ \gamma_\mu  \gamma_5
 \; , \; \slashed{A} \right\}  
\; , \; 
\tilde{S}_2
 \bar{R}^{\mu_2\nu_2} (x-y) 
\sigma_j
 \left[ \gamma_\rho \; , \; \gamma_{\mu_2} \right]
\right\}
 {j_A}_\nu^i (x,y)
\partial^\rho (j_{V,\nu_2}^j  (y,x) )
\nonumber
\\
&+&
\frac{i^4}{2}  e  ( \alpha g^2)^2
\; Tr \; 
 \left\{
\tilde{S}_2  \bar{R}^{\mu\nu} 
\hat{Q}
\sigma_i 
 \left\{ \gamma_\mu   \gamma_5
 \; , \; \slashed{A} \right\}  
\; , \; 
\tilde{S}_2
\bar{R}^{\mu_2\nu_2} (x-y) 
\sigma_j \gamma_5 
 \left\{ \gamma_\rho \; , \; \gamma_{\mu_2} \right\}
 \right\}
 {j_A}_\nu^i (x,y)
\partial^\rho    (j_{A,\nu_2}^j (y,x) ) 
,
\nonumber
\\
 I_{cross,B}&=&
-  \frac{i}{2}
(\alpha g^2)^2 e^2  
\;
 Tr 
\; 
\left[
\tilde{S}_2 
\left( 
\hat{Q}^2 (A_{\mu_2} A^{\mu_2} + A^{\mu_2} A^{\nu_2} \sigma_{\mu_2\nu_2})
+ \frac{i e}{2} \hat{Q}
 \sigma_{\mu_2\nu_2} F^{\mu_2\nu_2} \right)
\right.
\nonumber
\\
&\times& \left.
\tilde{S}_2 
 \bar{R}^{\mu\nu} \bar{R}^{\rho\sigma} 
\left( \gamma_\mu   
\gamma_\rho  \sigma_i  \sigma_j 
+
 \gamma_\rho
 \gamma_\mu \sigma_j  \sigma_i 
\right)\gamma_5 
\right]
 {j_V}_\nu^i (x,y) {j_A}_\sigma^j (y,x).
\end{eqnarray}
The following traces of isospin and Dirac indices ($Tr_F$ and $Tr_D$ )
 will be used in the next steps:
\begin{eqnarray}
Tr_F (\sigma_i \sigma_j) &=& 
2 \delta_{ij} ,
\\
Tr_F (\hat{Q} \sigma_i \sigma_j ) &=&
\frac{1}{3} \delta_{ij} + i \epsilon_{ij3},
\\
Tr_F ( \hat{Q}^2 \sigma_i \sigma_j ) &=&
\frac{5}{9} \delta_{ij} + \frac{i}{3} \epsilon_{ij3},
\\
Tr_D (\gamma^\mu \gamma^\nu) &=&
4 g^{\mu\nu} , 
\\
Tr_D (\sigma_{\mu\sigma} \sigma_{\rho\mu_2}) &=&
4 (
g_{\mu\mu_2} g_{\sigma\rho} 
- 
 g_{\mu\rho} g_{\sigma\mu_2} ) ,
\\
Tr_D
( \gamma^5 \gamma^\alpha \gamma^\beta \gamma^\delta \gamma^\lambda)
&=& - 4 i \epsilon^{\alpha \beta \delta \lambda}.
\end{eqnarray}
By resolving the effective coupling constants
in the long-wavelength limit, 
several of the terms above disappear.
Besides that, only the momentum derivatives of internal lines will be considered.
%i.e., the quark derivative couplings will be neglected.
The non zero contributions of these expressions can be written as:
\begin{eqnarray} \label{4quark}
{\cal L}_{4q} 
&=&  
\; 
\Delta_B M^* \bpsi \psi + 
  g_{4,B}  \left[ ( \bpsi  \psi )^2
 +   (  \bpsi \sigma_i  i \gamma_5 \psi )^2  \right]
+
\left(  \frac{3 g_{4,B}}{5} i \; \epsilon_{ij3}
+  g_{ps,B} \;c_i \; \delta_{ij} \right)
 (  \bpsi \sigma_i  i \gamma_5 \psi )
 (  \bpsi \sigma_j  i \gamma_5 \psi )
\nonumber
\\
&+&
\left[  \delta_{ij}  \left( g_{4v,B}  + g_{4v,B2}  \;c_i \;  
+ \frac{g_{4v,B-F}}{3}  
+ g_{4v2,B}
\right)
 + i  \epsilon_{ij3} 
\left( \frac{3}{5} g_{4v,B} 
+ g_{4v,B-F} 
+ 3  g_{4v2,B}
\right)
\right]
\nonumber
\\
&& \times
\left[ ( \bpsi  \sigma_i \gamma_\mu \psi ) ( \bpsi  \sigma_j \gamma^\mu \psi )
 + ( \bpsi \sigma_i \gamma_\mu \gamma_5 \psi )  
( \bpsi \sigma_j \gamma^\mu \gamma_5 \psi )
\right] 
\nonumber
\\
&+&
g_{s,sb}   ( \bpsi  \psi )^2 
+ 
g_{v,sb} ( \bpsi  \sigma_i \gamma_\mu \psi )^2
,
\end{eqnarray}
where
the following notation was adopted in the terms depending on the coefficients $c_i$
with operators $\Gamma_i$: 
\begin{eqnarray}
 \;c_i \; 
 (  \bpsi \Gamma_i \psi )^2  
=  \;c_1 \; 
 (  \bpsi  \Gamma_1 \psi )^2  
+ \;c_2 \; 
 (  \bpsi  \Gamma_2 \psi )^2 
+ 
\;c_3 \; 
 (  \bpsi  \Gamma_3 \psi )^2 ,
\end{eqnarray}
being defined 
the following isospin   coefficients:
\begin{eqnarray} \label{cj}
c_1=-\frac{4}{9}, \;\;\;\;\;\;\;
c_2 = \frac{4}{9},   \;\;\;\;\;\;\;
c_3 = \frac{5}{9} ,
\end{eqnarray}
These coefficients  are responsible  for the 
pseudo-scalar, vector and axial  quark-anti-quark states (pion, rho, $A_1$) 
 couplings
to the magnetic field.
In the first and second lines of expression 
(\ref{4quark}) there are  effective couplings dependent on 
the magnetic field and in the last line 
those due to the explicit symmetry breaking discussed in 
Ref. \cite{higher-orders}.
The couplings $g_{4,B}$ and mainly $g_{ps,B}$
  are responsible for extra contributions to the axial current and 
then they allow for chiral separation effect.
Chiral and isospin breaking terms  also  appear
 in the vector channel.
The effective coupling constants  are defined as:
\begin{eqnarray}
\Delta_B M^* &=&
- i  \; 2  e^2    (\alpha g^2) 
\; Tr \; \left[   
\hat{Q}^2
 \tilde{S}_2
(A_\mu A^\mu )
M^* \tilde{S}_2 R \right] 
,
\\
g_{4,B}  (1 ; (\delta_{ij} + \frac{3}{5}  i \epsilon_{ij3}\sigma_3) )
 &=&
- i  2 e^2   
 (\alpha g^2)^2
\; Tr \; \left[ 
\hat{Q}^2  \tilde{S}_2
A^\mu A_\mu
  \tilde{S}_2 R^2
\right]
 \left( 1 \; ; \; 
  \gamma_5^2     \sigma_i   \sigma_j
\right) ,
\\
g_{4v, B} (\delta_{ij} + \frac{3}{5} i \epsilon_{ij3}\sigma_3)  g^{\sigma_1\sigma}
 &=&    
- \frac{i}{2}    e^2   
(\alpha g^2)^2
\; Tr \; \left[  
\hat{Q}^2 
\sigma_i  \sigma_j  
 \gamma_{\rho_1}  \gamma_\rho 
\tilde{S}_2
 A^\mu A_\mu 
\tilde{S}_2
 \bar{R}^{\rho_1\sigma_1} \bar{R}^{\rho\sigma} 
\right]
,
\\
g_{ps,B}
\; c_i \delta_{ij} 
&=&  
-  i 2  e^2  (\alpha g^2)^2 
\; Tr \; 
\left[ \hat{Q} \sigma_i 
\hat{Q}  \sigma_j
\left(  \tilde{S}_2   
 \left[ \slashed{A} \; , \; 
\gamma_5      \right] R \right)
\left(
\tilde{S}_2 
  \left[ \slashed{A} \; , \; 
\gamma_5      \right]  R 
\right)
\right]
,
\\
g_{4v,B2} \; c_i \delta_{ij}  
 g^{\nu\sigma}
&=&
-\frac{i}{4} 
 e^2  ( \alpha g^2)^2  
\; Tr \;
\left[  \hat{Q}  \sigma_i \hat{Q}  \sigma_j 
\tilde{S}_2 
 \left[ \gamma_\mu \; , \; \slashed{A} \right] 
\bar{R}^{\mu\nu} 
\tilde{S}_2 
 \left[ \gamma_\rho \; , \; \slashed{A} \right] 
\bar{R}^{\rho\sigma}
\right]
,
\\
g_{4v,B-F}   \; (\delta_{ij} + 3 i \epsilon_{ij3}\sigma_3)
 g^{\nu\sigma}
&=&
\frac{i}{2} 
i e (\alpha g^2)^2
\; Tr \; \left[  \hat{Q} \sigma_i  \sigma_j 
F^{\mu_1\nu_1} 
\tilde{S}_2
\bar{R}^{\mu\nu}
\tilde{S}_2
 \bar{R}^{\rho\sigma}
\sigma_{\mu_1\nu_1}
 \gamma_\mu
 \gamma_\rho
\right],
\\
 g_{4v2,B} \;
(\delta_{ij} + 3 i \epsilon_{ij3}\sigma_3)
g_{\nu\nu_2}
&=&
- i e   ( \alpha g^2)^2   
 Tr \; 
\left[ 
\hat{Q} \sigma_i \sigma_j
(\partial_\sigma A_\rho) 
 \; [ \gamma^\mu \; , \; \gamma^\sigma ] 
[ \gamma^\rho \; , \; \gamma^{\mu_2} ]
\tilde{S}_2 \bar{R}_{\mu \nu} \tilde{S}_2 
\bar{R}_{\mu_2 \nu_2}
\right]
 ,
\\
g_{s,sb}  &=&
- i
4 (\alpha g^2)^2 
\; Tr \; 
({M^*} \tilde{S}_2 R)^2 ,
\\
g_{v,sb} \; \delta_{ij} g^{\nu\sigma}
&=&
-  i     (\alpha g^2)^2
\; Tr \; 
\left[
 \sigma_i  \sigma_j  \gamma_\mu   \gamma_\rho 
{M^*}  \tilde{S}_2 \bar{R}^{\mu\nu} \tilde{S}_2
{M^*}
 \bar{R}^{\rho\sigma}  
\right]
.
\end{eqnarray}
By performing the traces in discrete indices,  always
by neglecting the quark derivative couplings, 
and by taking $x = - i \frac{\partial}{\partial q_x}$
the above expressions 
can be written as:
\begin{eqnarray} \label{DBM}
\Delta_B M^* &=&
- i  \; \frac{40}{9} (e B_0)^2   
(\alpha g^2)  N_c
\; Tr' \;  
\left[   M^*
 \tilde{S}_2 
 x^2
\tilde{S}_2  
 R  
\right] 
, 
\\ \label{g4B}
g_{4,B} 
 &=&
- i  \; \frac{40}{9} (e B_0)^2  
 (\alpha g^2)^2 N_c
\; Tr' \;\left[   \tilde{S}_2  x^2 
 \tilde{S}_2  R^2  
\right] , 
\\ \label{g4vB}
g_{4v, B} g^{\sigma \sigma_1}
 &=&   
- i \frac{40}{9}   (e B_0)^2   
(\alpha g^2)^2 N_c
\; Tr' \; 
\left[  \tilde{S}_2 
x^2 
 \tilde{S}_2  
\bar{R}_2^{\sigma\sigma_1}  
\right] 
, 
\\ \label{gpsB}
g_{ps,B} c_j
&=&  
- i \;
 c_j  8  (e B_0)^2  (\alpha g^2)^2 N_c
\; Tr' \; 
\left[
( \tilde{S}_2 
x R )^2
\right]
,  
\\  \label{g4vB2}
g_{4v,B2} 
  g^{\nu\sigma} c_j
&=&
- i  \; 
 c_j 
 (e B_0)^2  ( \alpha g^2)^2  N_c
\; Tr' \; 
 \left[
\tilde{S}_2 
x
\bar{R}^{\mu\nu} 
\tilde{S}_2 
x 
\bar{R}^{\rho\sigma}
\right]
\left( 4 g_{\mu y}g_{y \rho} - 2 g_{\mu\rho} 
g_{yy}
\right)
,
\\   \label{g4vBF}
g_{4v,B-F} 
&=&
 i \frac{8}{3}  (e B_0) (\alpha g^2)^2 N_c
\; Tr' \; 
\left[ 
 \tilde{S}_2 \tilde{S}_2 
\bar{R}_2^{xy}
\right],
\\
\label{g4v2B-last}
 g_{4v2,B} 
  &=&
- i\; 16 (e B_0)  ( \alpha g^2)^2   \; N_c 
\;
 Tr' \;
\left[
T_{xy}
\left[ 
 8 
\tilde{S}_2  (R_T - R_L)
\tilde{S}_2  (R_T- R_L)
+ 
 \tilde{S}_2 (R_T-R_L)\tilde{S}_2 (R_T+R_L)
\right] 
\right]
 ,
\\ \label{gssb}
g_{s,sb}  &=&
- i 
 \;
32 (\alpha g^2)^2 N_c
\; Tr' \; 
({M^*} \tilde{S}_2 R)^2 ,
\\   \label{gvsb}
g_{
v,sb}  
g^{\nu\sigma} 
&=&
- i \;
8   (\alpha g^2)^2 N_c
\; Tr' \; 
\left[
{M^*}^2  \tilde{S}_2
\bar{R}^{\nu\mu}
\tilde{S}_2
 \bar{R}_{\mu}^{\sigma}  
\right]
,
\end{eqnarray}
where $T_{xy}= \frac{k_x k_y}{k^2}$
in expression (\ref{g4v2B-last})
and $Tr'$ stands for the trace/integral  in internal momenta.
Due to the structure of $\bar{R}^{xy}_2$ in expression (\ref{R2v})
the coupling  $g_{4v,B-F}$   is    non-zero 
only for a non-zero transversal  component of the gluon propagator, i.e.
if $R_T=0$ it yields $g_{4v,B-F}=0$.

\begin{figure}[ht!]
\centering
\includegraphics[width=140mm]{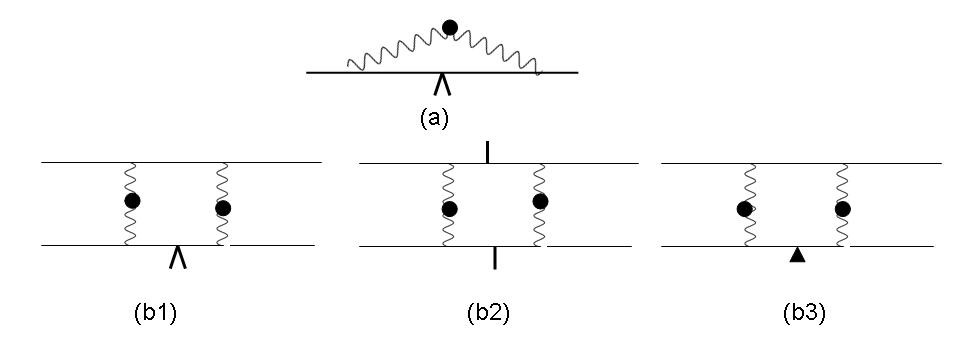}
\caption{ \label{diagrams}
In these diagrams, the wavy line with a full dot is a (dressed) non perturbative gluon propagator,
and the short bold lines insertions for the vector potential whereas 
the full triangle is for the magnetic field insertion ($F^{\mu\nu}$).
Diagram  (a) shows the effective mass due to the magnetic field contribution,
whereas diagrams (b1,b2,b3) represent all the quark-quark effective interactions 
shown above.
}
\end{figure}
  In Figure 1, the diagrams corresponding to the one loop terms
presented above are shown.
The wavy line with a full dot is a (dressed) non perturbative gluon propagator,
the short thick  lines insertions stand for the vector potential whereas 
the full triangle insertion  stands  for the magnetic field insertion ($F^{\mu\nu}$).
Diagram  (a) shows the contribution to the effective mass due to the magnetic field,
whereas diagrams (b1,b2,b3) represent  the quark-quark effective interactions 
shown above.

The  leading 
effective mass dependence on the magnetic field, 
shown in Fig. (1a), is of the order of 
 $(eB_0)^2/{M^*}^3$ 
 instead of the leading correction obtained from the gap equation $\sqrt{eB_0}$
\cite{Meff-B}.
The  leading coupling constants  in  the expressions above
are  $g_{4v2,B}$
and  $g_{4v,B-F}$
that are linearly 
proportional to the magnetic field $\partial_\mu A_\nu$,
 $eB_0/{M^*}^2$.
 The corresponding diagram is shown in Fig. (1b3). 
By extracting $1/{M^*}^2$ from $\tilde{S}_2$ in the limit of large 
quark effective mass, it produces a quantity proportional to 
the dipole moment coupling  itself $eB_0/(2M^*) $.
 In spite of the absence
of a tensor current for the  dipolar coupling  in the leading 
effective action, the magnetic field couples directly to the vector/axial currents
being a dipolar interaction.
%{\bf  ***
% The emergence of an explicit magnetic moment effective Lagrangian 
%term  is envisaged  for flavorless fermions in  Ref. \cite{magmom}.
%}
All the other couplings - in expressions (35)-(39) - 
  have two insertions $A_\mu A^\mu$ introducing a larger (and suppressing)
momentum dependence in internal lines, with corresponding factor
$(eB_0)^2/(M^*)^4$.
 They  are   smaller in the limit of large quark effective mass.

\section{Ratios between effective coupling constants}
\label{sec:numerical}

 There are few  ambiguities in performing numerical estimates of the 
effective coupling constants found above. 
The first reason is that 
the gluon propagator with its
infrared behavior is not really well known and results  depend strongly on it.
Also, one has to choose a way of performing  the
momenta/energy traces, 
for example  in Euclidean or Minkowski spaces,
and this might yield different numerical results.
Furthermore, 
other effects in the gluon sector, such as the $B_0$ dependence of the 
quark gluon coupling or the gluon propagator itself,   might be expected to yield
$B$-dependence at least of the same order of magnitude as the  quark condensate (or quark
effective mass) from the gap equation \cite{bali-etal1303.1328}.
Due to these reasons numerical estimates will not be presented.
Nevertheless, below few solutions for the gap equation are presented
with the intention to justify the approximations done, i.e.
to consider the quark effective mass from 
DChSB as the leading effect of the auxiliary field and 
the large quark effective mass expansion.
With respect to the gap equation, 
the behavior of the chiral condensate, and therefore of the quark effective mass, 
 under a constant magnetic field 
has been investigated extensively 
 \cite{review-B-general,ritus,bali-etal1303.1328} and it has been found that
the increase of the quark condensate with the (weak) magnetic field
is due  to the increase of the density of states  by accounting 
the lowest Landau levels with high degeneracy in this regime.
Besides that,  magnetic catalysis has also been related to the positivity
of the scalar QED $\beta-$ function \cite{cond-B}.

To solve the gap equation (\ref{gap-eq})
a  longitudinal (confining) effective gluon propagator was chosen   of the
form of: 
$g^2 R^{\mu\nu}_{ab} (k) = K_F/(k^2 + M_g^2)^2 g^{\mu\nu} \delta_{ab} $ 
for $K_F= 8\pi^3 M^2/9$
\cite{greensite,cornwall}. 
This effective propagator incorporates the large strength of the 
running coupling constant and, to some extent, 
some of the relatively important issues of the ultraviolet and  infrared behavior 
of the gluon confining propagator \cite{cornwall}.
 With this gluon propagator the gap equation, as well as all the 
expressions for the effective quark masses and effective coupling
constants, are finite, i.e. free of ultraviolet and 
infrared divergences.
For example consider $M_g = M \simeq 378$ MeV, that is 
of the order of the values discussed in Ref. \cite{cornwall} in spite of being relatively smaller 
than the usual  theoretical and lattice  findings \cite{gluonmass,SD-rainbow}.
For a current quark mass $m=10$MeV,
the gap equation (\ref{gap-eq}) 
is non zero only for the scalar auxiliary field $s$ as defined,
and it  yields, for $B_0=0$, $s_{0} \simeq 210$ MeV for which 
$M^* \simeq  220$ MeV. 
For a weak magnetic field 
$z=\frac{e B_0}{{M^*}^2} = 0.1$ 
the gap equation yields 
$ M^* (z=0.1) = 227$ {MeV}.
By considering   $M_G= 511$ MeV, which is closer to the values obtained in lattice  QCD
it yields 
$M^* (z=0) = 300$ {MeV} and for weak magnetic field
$M^* (z=0.1) = 309$ {MeV}.

The effective coupling constants presented above
can exhibit simple relations in the limit of large effective masses.
For some of these effective coupling constants, 
  this is achieved in  specific  limits of the 
 gluon kernels.
In  the limit of very large quark effective mass,
i.e. for $\tilde{S}_2 \to 1/{M^*}^2$,
 some of these ratios are independent 
 of the chosen component for the 
gluon propagator ($R_T(x-y)$ or $R_L(x-y)$),
i.e. :
\begin{eqnarray}
\frac{\Delta M^*_B}{\Delta M^*} &\sim& \frac{5}{9} \frac{(e B_0)^2}{ {M^*}^4}
,
\;\;\;\;\;\;\;
\frac{g_{4,B}}{g_4} \sim
\frac{5}{9}  \frac{(e B_0)^2}{ {M^*}^4}
,
\;\;\;\;\;\;\;
\frac{g_{ps,B}}{g_{4,B}} \sim
\frac{9}{5}
,
\;\;\;\;\;\;
\frac{g_{4v,B2}}{g_{4v,B}} 
\sim 
\frac{27}{40}
\end{eqnarray}

For other effective coupling constants,
still in the limit of very large effective quark mass, $M^*$ ,
 it is possible to  obtain simple relations 
by considering particular  relative contributions  of the longitudinal and transversal
components of the gluon propagator.
In the following it   will be considered that 
any  of the two  components  present an effective gluon mass.
The ratios will be computed
 in the limit of large masses $R_{T/L} \sim 1/M_G^n$ for $n = 2, 4$,
 by keeping $M_G > M^*$.
If 
it is  assumed  
$R_L=0$  then the following ratios are obtained:
\begin{eqnarray}
\left( \frac{g_{4,B}}{g_{4v,B}} \right)^T
 \sim \frac{3}{4} ,
\;\;\;\;\;
\left(
\frac{g_{4,B-F}} {g_{4,B}}
\right)^T
\sim 
\frac{3}{20} \frac{{M^*}^2}{e B_0},
\;\;\;\;\;
\left( \frac{g_{4,B}}{g_{4v2,B}} 
\right)^T
\sim 
\frac{80}{9} \frac{e B_0}{{M^*}^2},
\end{eqnarray}
whereas for $R_T=0$
 it yields:
\begin{eqnarray}
\left( \frac{g_{4,B}}{g_{4v,B}} 
\right)^L
\sim  \frac{1}{4}
 , 
\;\;\;\;\;
\left( \frac{g_{4,B-F}}{g_{4,B}} 
\right)^L
\sim  0,
\;\;\;\;\;
\left(
\frac{g_{4,B}}{g_{4v2,B}} 
\right)^L
\sim 
\frac{80}{63} 
 \frac{e B_0}{{M^*}^2}
 .
\end{eqnarray}
All the coupling constants of the order of $B_0$ or $B^2_0$ 
are smaller than the NJL coupling, from expression (\ref{NJL-couplings}),
since a large effective quark mass expansion has been done:
i.e.
$\frac{(eB_0)}{{M^*}^2} < 1$  or $\frac{(eB_0)}{{M^*}^2} << 1$.
However, 
by increasing the magnetic field strength 
this expansion still may be reliable up to some limit  
 by computing
 higher orders terms ($n-$th order expansion). 
This produces further quark-quark effective 
 interactions dependent on $B_0^{n-j}$
where $j=0,1,2...n$.
Consequently the complete account of the 
Landau orbits that could be done for the quark kernel
\cite{ritus,review-B-general} emerges as a series in powers of the magnetic field 
in agreement with \cite{weak-B}.

\section{ Summary and Conclusions}

By departing from a (dressed) one gluon exchange mechanism for the quark-quark interaction,
different leading quark-quark effective interactions due to polarization
 were derived in the presence of a 
weak magnetic field,
%with respect to the quark effective mass,
 i.e. 
$e B_0 < < {M^*}^2$.
 The relevant assumption for the GCM is that the gluon propagator 
is dressed by non perturbative effects due to
 the non Abelian character of gluon interactions.
The one loop BFM method was applied with a
correction due to the auxiliary field method. 
However only the  leading effect of the auxiliary fields was considered,
that is the correction to the quark effective mass.
The one-loop quark  effective action in the presence of the background 
field was expanded  
for
 large quark effective mass and 
weak magnetic field
up to the second order  in  
quark bilinears and to leading order
 in the  magnetic field.
The (leading) first and second order $B_0$-dependent terms provided corrections to
the background quark mass and effective interactions 
such as  the usual NJL and vector NJL ones, besides new 
   chiral and isospin symmetry breaking terms.
They  correspond to the different couplings of 
the magnetic field to 
pseudo-scalar, vector and axial isospin triplets states.
The set of $B_0$-dependent  interactions from expressions
(\ref{4quark})
 is given by:
\begin{eqnarray} \label{4quark-final}
{\cal L}_{4q} 
&=&  
\; 
\Delta_B M^* \bpsi \psi + 
  g_{4,B}  \left[ ( \bpsi  \psi )^2
 +   (  \bpsi \sigma_i  i \gamma_5 \psi )^2  \right]
+
\left(  \frac{3 g_{4,B}}{5} i \; \epsilon_{ij3}
+  g_{ps,B} \;c_i \; \delta_{ij} \right)
 (  \bpsi \sigma_i  i \gamma_5 \psi )
 (  \bpsi \sigma_j  i \gamma_5 \psi )
\nonumber
\\
&+&
\left[  \delta_{ij}  \left( g_{4v,B}  + g_{4v,B2}  \;c_i \;  
+ \frac{g_{4v,B-F}}{3}  
+ g_{4v2,B}
\right)
 + i  \epsilon_{ij3} 
\left( \frac{3}{5} g_{4v,B} 
+ g_{4v,B-F} 
+ 3  g_{4v2,B}
\right)
\right]
\nonumber
\\
&& \times
\left[ ( \bpsi  \sigma_i \gamma_\mu \psi ) ( \bpsi  \sigma_j \gamma^\mu \psi )
 + ( \bpsi \sigma_i \gamma_\mu \gamma_5 \psi )  
( \bpsi \sigma_j \gamma^\mu \gamma_5 \psi )
\right] 
\nonumber
\\
&+&
g_{s,sb}   ( \bpsi  \psi )^2 
+ 
g_{v,sb} ( \bpsi  \sigma_i \gamma_\mu \psi )^2
,
\end{eqnarray}
The mass correction $\Delta_B M$ 
 is positive as expected from the usual 
magnetic catalysis analysis from the NJL-type gap equation.
However this effective mass contribution for weak field is of the order 
of $(eB_0)^2/{M^*}^3$
whereas
 the leading contribution from the gap equation for weak $B$-field
is of the order of $\sqrt{eB_0}$.
The gap equation for the auxiliary field  was found to depend on the magnetic 
field as usually investigated for NJL or GCM -type models.
Almost all the effective coupling constants are 
of the order of 
$(e B_0)^2/{M^*}^4$
except two of them, $g_{4v2,B}$ and $g_{4v,B-F}$
are  ${\cal O} (e B_0/{M^*}^2)$, corresponding 
therefore to dipolar couplings  in spite of 
the absence of the tensor current.
 These two  effective couplings are the leading ones being that
 $g_{4v,B-F}$ is non zero only if the gluon propagator has a transversal component.
There are  overall corrections 
to the NJL and vector-NJL coupling constants
respectively given by: $g_{4,B}$ and 
  $g_{4v,B}$, $g_{4v,B-F}$  and $g_{4v2,B}$.
The effective coupling constant $g_{4,B}$  enhances the
strength of the quark scalar interaction
This might be seen as an increase of the 
strength of quark interactions that produce
  dynamical chiral symmetry breaking.
Although this may suggest that DChSB can be obtained for zero NJL coupling constant
($g_4 \to 0$)
when $g_{4,B} \neq 0$,  this might be misleading in the sense that 
 in the present development  both effective couplings have the same physical 
origin, namely the quark  polarization with a quark-gluon coupling  $g^2$.
The physical content of magnetic catalysis would be clearer
in this sense  by considering
a different mechanism for one of the two effective 
interactions ($g_4$ or $g_{4,B}$).
The effective coupling $g_{4v2,B}$ is also positive
and therefore it might  contribute to the vector condensation in the
vacuum \cite{chernodub}.
However some new couplings appear signaling the emergence of 
pseudo-scalar and vector/axial multiplets with different
interaction with the magnetic field, i.e. different  electric charge
($+$, $-$ and $0$).
These couplings therefore break chiral and isospin symmetries.
In particular the effective couplings  $g_{4,B}$ and $g_{ps,B}$  yield
 pions interactions with the magnetic field.
The vector couplings to the magnetic field are
$g_{4v,B}$, $g_{4v,B2}$, $g_{4v,B-F}$ and $g_{4v2,B}$
providing the different couplings in the vector and axial channels
therefore related to the $\rho$ and $A_1$ triplets.
The two couplings due to the explicit symmetry breaking,
$g_{s,sb}$ and $g_{v,sb}$,
 have already
been  investigated in Ref. \cite{higher-orders}.
The analytical ratios exhibited in Section \ref{sec:numerical} are very specific to the 
limit in which the large quark effective mass is smaller than an effective gluon mass
that is expected to be present in a non perturbative gluon propagator 
\cite{gluonmass}.
Other limits could be considered and will be presented elsewhere.
The main sources of possible  improvements are 
 the simplified  momentum dependence of the internal lines and
the inclusion of  auxiliary fields which however will produce numerically smaller contributions.
Higher order interactions of
 the expansion of the quark determinant considered in this manuscript
yield corrections for stronger magnetic fields with increasing
powers of  $eB_0$ 
for the  effective coupling constants. 
Alternatively, 
the whole summation over the Landau levels 
for  internal quark lines (quark kernel) can be considered  for arbitrary values of the 
magnetic field.
Pion and quark $B_0$-dependent effective interactions 
 will be investigated elsewhere.

\section*{Acknowledgments}

The author thanks short conversation with I. Shovkovy 
 and
 acknowledges partial financial support from CNPq.

\end{document}